\documentclass[runningheads]{llncs}
\usepackage{cite}
\usepackage{amsmath,amssymb,amsfonts}
\usepackage{algorithmic}
\usepackage{graphicx}
\usepackage{textcomp}
\usepackage{xcolor}

\usepackage{balance}
\usepackage{booktabs}
\usepackage{multirow}
\usepackage{mathtools}
\usepackage{bm}
\usepackage[normalem]{ulem} 
\usepackage{hyperref}

\usepackage{courier}
\usepackage[utf8]{inputenc}
\usepackage{caption}
\usepackage{nicematrix}
\usepackage{fontawesome5}
\begin{document}
\title{Learning Audio–Visual Embeddings with Inferred Latent Interaction Graphs}
\author{Anonymous}
\institute{}
%
\author{Donghuo Zeng\thanks{Corresponding Author} \orcidID{0000-0002-6425-6270} \and
Hao Niu \and
Yanan Wang \and
Masato Taya }
\authorrunning{Zeng et al.}
\institute{KDDI Research, Inc., Saitama, Japan \\ \email{\{xdo-zen, ha-niu, wa-yanan, ma-taya\}@kddi.com}}
\maketitle              
\begin{abstract}
Learning robust audio–visual embeddings requires bringing genuinely related audio and visual signals together while filtering out incidental co-occurrences—background noise, unrelated elements, or unannotated events. Most contrastive and triplet-loss methods use sparse annotated labels per clip and treat any co-occurrence as semantic similarity. For example, a video labeled “train” might also contain motorcycle audio and visual, because “motorcycle” is not the chosen annotation; standard methods treat these co-occurrences as negatives to true motorcycle anchors elsewhere, creating false negatives and missing true cross-modal dependencies.
We propose a framework that leverages soft-label predictions and inferred latent interactions to address these issues:
(1) \textit{Audio–Visual Semantic Alignment Loss (AV-SAL)} trains a teacher network to produce aligned soft-label distributions across modalities, assigning nonzero probability to co-occurring but unannotated events and enriching the supervision signal.
(2) \textit{Inferred Latent Interaction Graph (ILI)} applies the GRaSP algorithm to teacher soft labels to infer a sparse, directed dependency graph among classes. This graph highlights directional dependencies (e.g., “Train (visual)” $\rightarrow$ “Motorcycle (audio)”) that expose likely semantic or conditional relationships between classes; these are interpreted as estimated dependency patterns.
(3) \textit{Latent Interaction Regularizer (LIR)}: A student network is trained with both metric loss and a regularizer guided by the ILI graph, pulling together embeddings of dependency-linked but unlabeled pairs in proportion to their soft-label probabilities.
Experiments on AVE and VEGAS benchmarks show consistent improvements in mean average precision (MAP), demonstrating that integrating inferred latent interactions into embedding learning enhances robustness and semantic coherence.

\end{abstract}
\section{Introduction}
Learning aligned representations across audio and visual modalities is fundamental to tasks such as cross-modal retrieval~\cite{zeng2018audio,Suris2018, zeng2019learning} and audio-visual event localization~\cite{tian2018audio}. The standard approach projects audio and visual inputs into a shared embedding space via metric learning (e.g., contrastive or triplet loss), pulling annotated positive pairs together and pushing negatives apart.

However, real‐world audio–visual data are far richer than the sparse annotations typically provided. Datasets often assign only one audio label and one visual label per clip, even when multiple events occur simultaneously.
For instance, a short video clip might be annotated “train” for both audio (the sound of a train horn or engine) and visual (the appearance of a train) tracks, yet that same clip can also contain a motorcycle’s engine revving and a passing car’s headlights. These unannotated audio and visual elements coexist with the annotated content, contributing to the true semantics of the scene—even if they are deemed “not important” during annotation. Unfortunately, when a model trains with conventional losses on such data, it treats these unlabeled co‐occurrences that lack matching labels as negatives. Concretely, if “train” is the only annotated label, then the unannotated “motorcycle” audio and “motorcycle” visual—even though they appear together—are considered unrelated, so their embeddings are pushed apart. This introduces two fundamental issues: (I)~\textbf{False negatives due to annotation incompleteness.} By relying solely on annotated labels, standard metric‐learning methods unavoidably treat semantically meaningful but unlabeled co‐occurring audio–visual pairs as negatives~\cite{hadsell2006dimensionality, khosla2020supervised, hermans2017defense, zeng2020deep, zeng2022complete}. Consequently, the model learns to repel embeddings that are actually related in real‐world scenes. The effect is that “motorcycle” audio and “motorcycle” visual—both present in the same "Train" clip—are mistakenly regarded as dissimilar, degrading the retrieval of genuinely related pairs. 
(II)~\textbf{Spurious correlations from incidental co‐occurrences}. Beyond simply overlooking unlabeled events, standard methods~\cite{radford2021learning, li2019visualbert, lu2019vilbert} also fail to distinguish between truly semantic co‐occurrences and incidental or background co‐occurrences~\cite{kim2023exposing, agarwal2020towards}. In noisy or complex scenes, overlapping elements are common; for example, a train scene may include traffic, wind, or construction sounds. Without mechanisms to filter out such incidental signals, models trained solely on labeled data may incorrectly associate background audio (e.g., traffic hum) with a target event (e.g., a train) due to frequent co‐occurrence.  This leads to spurious correlations that hurt generalization: At test time, a model may retrieve a "Train (visual)" in response to traffic sounds alone.

Together, these issues mean that existing metric-learning~\cite{zeng2024anchor} approaches produce embeddings that neither capture the full richness of real‐world audio–visual co‐occurrences nor effectively filter out background noise. In other words, two clips that genuinely share a latent relationship, such as a clip containing both "train" and "motorcycle", can be misrepresented as unrelated, while two clips that merely share background context, such as traffic noise and train passing, can be incorrectly treated as strongly related. Our key insight is that correlations differ in importance: some reflect meaningful semantic dependencies, while others are incidental. 

To address this, we introduce a framework that leverages predicted soft-label distributions and dependency inference to improve embedding learning:
(1)~\textbf{AV-SAL} trains a \emph{teacher} network with a semantic-alignment loss to produce soft labels across all classes, capturing latent events beyond sparse annotations.
(2)~\textbf{ILI Graph}: 
We apply the GRaSP~\cite{lam2022greedy} algorithm to the teacher’s soft labels to infer a sparse, directed dependency graph over classes. The graph encodes asymmetric predictive dependencies in the soft-label/representation space; while these patterns may be interpretable under standard structural assumptions, we treat them as an estimated dependency structure from observational model outputs—not definitive causal proof.

(3)~\textbf{LIR}: We train a \emph{student} network with a latent interaction regularizer derived from the ILI graph, ensuring that both annotated and dependency-linked unlabeled pairs reinforce the representation, while incidental pairs are down-weighted.

We validate our approach on the AVE and VEGAS benchmarks, where multi-event co-occurrences are common. Our method consistently outperforms state-of-the-art baselines by approximately 1.5\% in mean average precision (MAP), demonstrating that latent interaction inferring significantly enhances the robustness and semantic coherence of audio–visual embeddings. 

\section{Related Work}
\subsection{Audio–Visual Embedding Learning}
Learning aligned representations across audio and visual modalities has been widely studied using linear, nonlinear, and deep-network approaches. Early work relied on Canonical Correlation Analysis (\textit{CCA})~\cite{hardoon2004canonical} and its extensions (\textit{Cluster CCA}~\cite{rasiwasia2014cluster}, \textit{K-CCA}~\cite{akaho2006kernel}), which find correlated projections but struggle with nonlinear relationships and scale poorly. Deep CCA variants (\textit{DCCA}~\cite{andrew2013deep}, \textit{C-DCCA}~\cite{yu2018category}, \textit{VAE-CCA}~\cite{zhang2023variational}) leverage neural networks to model nonlinearity and within-class correlation but remain constrained by sparse annotations and ignoring unlabeled co-occurrences. Deep-network–based methods extract embeddings from supervised classifiers~\cite{roth2020ava, torfi20173d, cheng20look, zheng2021deep, tian2018audio} or self-supervised correspondence objectives~\cite{arandjelovic2017look, korbar2018cooperative}, capturing rich co-occurrence patterns. While effective, these approaches do not explicitly separate true semantic dependencies from incidental or background co-occurrences. Dual-subspace approaches~\cite{zeng2023learning} attempt to separate shared and modality-specific features but still rely on labeled data for subspace formation.

Metric-learning approaches~\cite{hadsell2006dimensionality, hermans2017defense, zeng2022complete, zeng2020deep, zeng2024anchor} enforce embedding distances directly via contrastive or triplet losses. Distillation-based methods~\cite{wang2023videoadviser, zeng2025metric} further leverage teacher networks and soft-label distributions to enrich supervision. Despite their effectiveness, these methods still treat unlabeled but meaningful co-occurrences as negatives, leading to false negatives and spurious correlations.

Overall, existing audio–visual embedding approaches remain limited by sparse annotations, lack mechanisms to capture latent events, and cannot dynamically differentiate incidental co-occurrences from semantically meaningful dependencies. This motivates frameworks that combine soft-label supervision with dependency inference to better reflect the richness of real-world scenes.

\subsection{Latent Interaction and Dependency Learning}
Instead of relying solely on correlation-based embeddings, recent works have explored learning latent interaction structures to capture meaningful dependencies among modalities~\cite{pearl2018book}. These approaches aim to infer underlying dependency patterns from observational data~\cite{pearl2018book, hogan2019causal, wang2019blessings}. In high-level vision and language tasks (e.g., VQA~\cite{Teney_2021_ICCV}, image captioning~\cite{yang2021deconfounded}, visual dialogue~\cite{qi2020two}), modeling latent interactions has been shown to reduce spurious correlations, disentangle confounding factors, and improve generalization. Similarly, in audio–visual learning, capturing such dependencies can mitigate the impact of missing annotations and incidental co-occurrences.  Beyond general dependency inference, structured graphs have been used to explicitly encode relationships among classes or embeddings. For instance, GRaSP~\cite{lam2022greedy} constructs sparse directed dependency graphs from soft-label predictions by permutation search with BIC scoring. Such graphs highlight consistent interaction patterns without requiring ground-truth causation and can guide embedding learning by reinforcing semantically linked pairs while down-weighting incidental ones, enabling models to dynamically uncover latent semantic structures under sparse or noisy annotations.

By integrating soft-label supervision with latent interaction inference, models can move beyond static correlation-based embeddings, uncover deeper semantic dependencies, and produce more explainable and transferable audio-visual representations~\cite{zeng2025metric, zhang2024causal}.

\section{Proposed Method}
\subsection{Notation and overview}
Let $\mathcal{D}=\{x_n\}_{n=1}^N$ be a dataset of $N$ clips. Each clip $x_n$ has audio-visual pair $(x_{a,n},x_{v,n})$.  
The teacher model produces \textbf{soft-label} distribution $s_n \in [0,1]^C$ (independent sigmoid heads per class, $C$ is the number of classes); collect these into $S\in\mathbb{R}^{N\times C}$.  
The student embedding network is $z_\theta(x)\in\mathbb{R}^d$ and $D(z,z')$ denotes the Euclidean distance.

Our pipeline follows a two-stage schedule (teacher $\rightarrow$ student). After training the teacher until epoch $E_{\text{teacher}}$, we collect $S$ and infer a sparse, directed inferred latent interaction (ILI) adjacency $\widehat{A}=\mathrm{GRaSP}(S)\in\mathbb{R}_{\ge0}^{2C\times 2C}$. $\widehat{A}_{ij}$ quantifies a discovered dependency from class $i$ to class $j$ ($i,j \in \{1,2, ..., 2C\}$). During student training, we augment the metric objective with a Latent Interaction Regularizer (LIR) weighted by $\widehat{A}$.

\begin{figure}[t]
  \centering
  \includegraphics[width=0.95\linewidth]{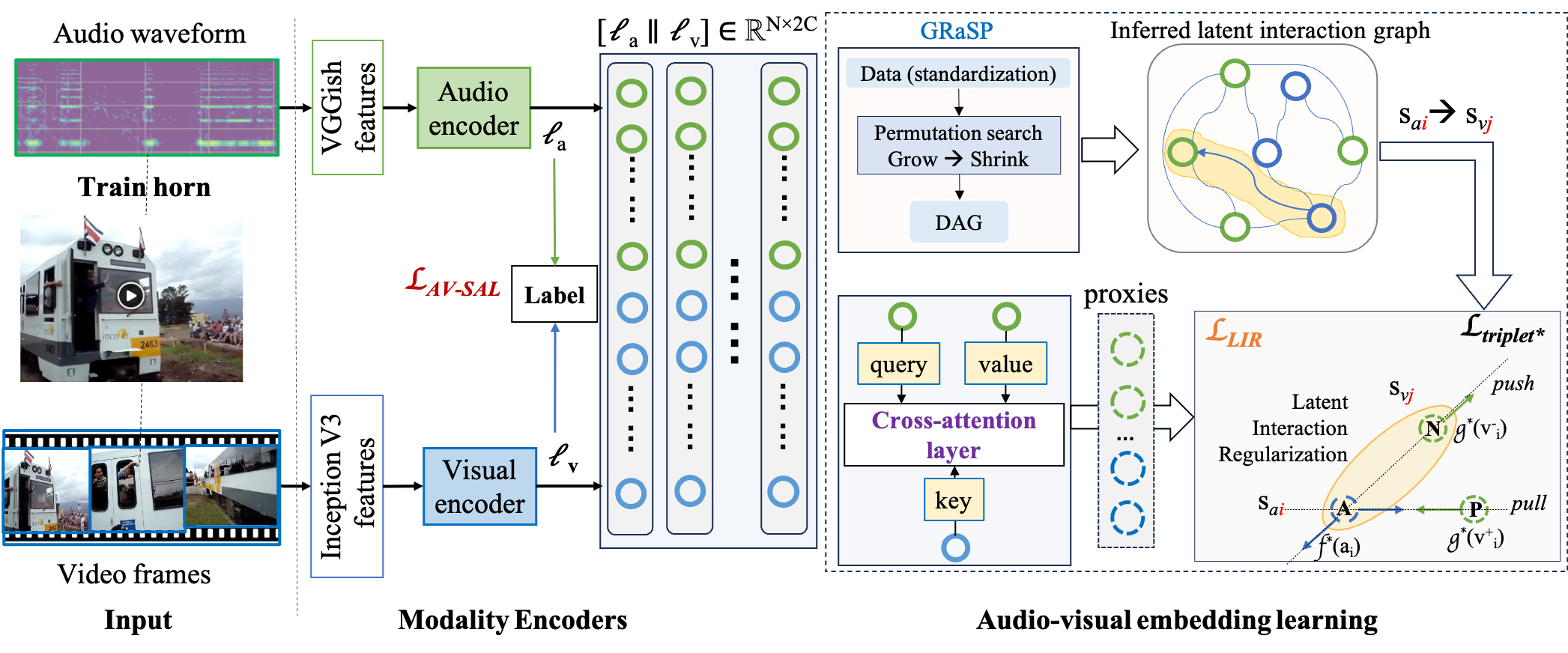}
  \caption{\textbf{Overview of our architecture.} Audio and visual features are extracted using VGGish and Inception V3. Logits from the audio and visual encoders are optimized with AV-SAL combined with a proxy-triplet loss (incorporating cross-modal attention) to produce calibrated soft-label distributions. These soft labels are input to GRaSP to infer a directed latent interaction graph $\widehat{A}$, where each edge $s_{a,i} \rightarrow s_{v,j}$ represents a dependency between audio label $i$ and visual label $j$. Finally, the embeddings are refined with a LIR term—weighted and normalized according to the discovered dependencies.}
  \label{fig:arch}
\end{figure}
\subsection{Teacher: audio–visual semantic alignment (AV-SAL)}
The teacher model is trained to produce calibrated soft-labels that \emph{discover} latent co-occurrences beyond sparse annotations. We use an alignment loss between modality soft-labels $s_n$ and available one-hot labels $Y_n\in\{0,1\}^C$:
\begin{equation}
\mathcal{L}_{\mathrm{AV\text{-}SAL}} \;=\; \frac{1}{N}\sum_{n=1}^N \Big(\|s_{a,n}-Y_n\|_2^2 + \|s_{v,n}-Y_n\|_2^2\Big),
\end{equation}
\small
where $s_{a,n},s_{v,n}\in[0,1]^C$ are the teacher's per-modality sigmoid outputs for clip $n$. The teacher is also trained with a proxy-triplet~\cite{zeng2024anchor} loss, so that soft labels reflect both annotation and embedding:
\begin{equation}
\small
\mathcal{L}^{\mathrm{(teacher)}}=\mathcal{L}_{\mathrm{AV\text{-}SAL}}+\mathcal{L}_{\mathrm{triplet*}}.
\end{equation}
Given proxy embeddings \(\ell_{a*}^{(i)}\) and \(\ell_{v*}^{(i)}\) extracted from a cross-attention layer~\cite{zeng2024anchor}, the proxy triplet loss is defined as
\begin{equation}
\small
\label{eq:triplet_proxy}
L_{\text{triplet}^*}
=\frac{1}{N}\sum_{i=1}^N
\Big[ \|\ell_{a*}^{(i)}-\ell_{v*}^{(i+)}\|_2^2
- \|\ell_{a*}^{(i)}-\ell_{v*}^{(i-)})\|_2^2
+\alpha \Big]_+,
\end{equation}
where $\alpha$ is a margin hyperparameter that enforces the anchor to be closer to the positive than the negative by at least $\alpha$. Here,  $(\ell_{a*}^{(i)}, \ell_{v*}^{(i+)})$ form a positive pair (same label), while $\ell_{v*}^{(i-)}$ is sampled from a different class.

\subsection{Inferring the ILI Graph}
After the teacher stage (epoch \(E_{\mathrm{teacher}}\)) we collect the teacher \emph{logits} for all clips and form the logits data matrix:
\[
\mathcal{L} = \bigl[\ell^{(1)},\dots,\ell^{(N)}\bigr]^\top
\;\in\;\mathbb{R}^{N\times 2C},
\]
where each row \(\ell^{(i)}=[\ell_a^{(i)},\ell_v^{(i)}]\) concatenates the \(C\) audio and \(C\) visual logits for clip \(i\). Because logits may have different scales across classes, we column-wise standardize \(\mathcal{L}\) to obtain \(\widetilde{\mathcal{L}}\) (zero mean and unit variance per column). We apply GRaSP \cite{lam2022greedy} to \(\widetilde{\mathcal{L}}\). GRaSP searches permutation (variable-ordering) space via a greedy tuck/score routine and, for each candidate ordering \(\pi\), induces per-node parent sets \(P_j(\pi)\) using the Grow–Shrink procedure. The canonical GRaSP local score is BIC for a Gaussian linear model; GRaSP selects the ordering \(\pi^\star\) that minimizes the total score and returns the induced sparse DAG.

As a robustness-oriented, application-specific adaptation, for each node \(j\) with predecessors \(P_j(\pi)\) we optionally fit a nodewise \(\ell_1\)-regularized linear model on the standardized logits:
\begin{equation}
\small
\widehat{\beta}^{(j)}(P_j(\pi)) \;=\; 
\arg\min_{\beta}\;
\frac{1}{2n}\big\|\widetilde{\mathcal{L}}_{:,j}-\widetilde{\mathcal{L}}_{:,P_j(\pi)}\beta\big\|_2^2
\;+\;\lambda_{\mathrm{reg}}\|\beta\|_1,
\label{eq:nodewise-lasso}
\end{equation}
where \(\lambda_{\mathrm{reg}}\ge0\) is a small regularization parameter (setting \(\lambda_{\mathrm{reg}}=0\) reduces \eqref{eq:nodewise-lasso} to ordinary least squares). We take preliminary directed weights
\begin{equation}
\small
\widetilde{A}_{i\to j} \;=\;
\begin{cases}
\big|\widehat{\beta}^{(j)}_i\big|, & i\in P_j(\pi^\star),\\[2pt]
0, & \text{otherwise},
\end{cases}
\end{equation}
i.e., coefficients are zero for non-parents. GRaSP selects the best ordering \(\pi^\star\) and the corresponding parent sets \(P_j(\pi^\star)\) via its greedy tuck/score procedure; the nodewise fits above are used only to provide coefficient-based weights on the selected edges. To form the final adjacency used by LIR, we (i) set the diagonal entries of \(\widetilde{A}\) to zero and (ii) normalize the matrix by its \(L_1\)-norm:
\begin{equation}
\small
\widetilde{A}_{ii} := 0\quad\forall i,\qquad
\widehat{A} \;=\; \frac{\widetilde{A}}{\|\widetilde{A}\|_1}.
\label{eq:normalize-A}
\end{equation}
The resulting \(\widehat{A}\in\mathbb{R}_{\ge0}^{2C\times 2C}\) is a sparse, directed dependency matrix (the ILI graph) that weights LIR (Eq.~\eqref{eq:LIR}). As a practical note, interpret \(\widehat{A}\) as an estimated dependency graph derived from teacher logits rather than definitive causal evidence; accordingly, we report edge stability (via bootstrap / different seeds) and restrict LIR to high-weight, stable edges in experiments. The \(\ell_1\)-regularized nodewise regressions and the diagonal/normalization in \eqref{eq:normalize-A} are implementation choices made for robustness and interpretability.

\subsection{Latent Interaction Regularizer (LIR)}
LIR uses $\widehat{A}$ to pull together student embeddings for clips predicted to co-activate dependent classes, thereby correcting false negatives that arise from sparse annotations. Define a sampling distribution $\mathcal{S}_{ij}$ that draws audio/visual clip pairs $(n,m)$ for which class $i$ (resp.\ $j$) is active in the teacher soft-labels (e.g., $s_n[i]>\tau$). The LIR is
\begin{equation}
\small
\mathcal{L}_{\mathrm{LIR}}(\theta;\widehat{A})
\;=\; \sum_{i=1}^C\sum_{j=1}^C W((m, i), (n, j))\widehat{A}_{ij}\;\mathbb{E}_{(n,m)\sim\mathcal{S}_{ij}}\big[\,D\big(z_\theta(x_n),\,z_\theta(x_m)\big)\,\big].
\label{eq:LIR}
\end{equation}
In practice, we approximate the expectation by mini-batch sampling and normalize $\widehat{A}$ so that $\sum_{i,j}\widehat{A}_{ij}=1$. We set $W(a, a) = W(v, v)$ =0.1, and $W(a, v) = W(v, a)$ =0.4, emphasizing cross-modal latent interaction links.

\subsubsection{Overall training objective}  
Both teacher and student networks share the same modality encoders and embedding architecture. We adopt a two‐stage schedule with transition epoch $M\in\{300,400,500,\dots,900\}$. The goal of Teacher stage (1 to $M$) is to learn well-calibrated soft-labels for latent discovery. In Student stage ($M$ to 1000), after inferring the latent interaction graph, we incorporate LIR to guide embedding learning with causal structure:
\begin{equation}
    \small
    \mathcal{L}^{(student)} \;=\; \mathcal{L}_{\text{AV-SAL}} \;+\; \mathcal{L}_{\text{triplet*}} \;+\; \gamma\,\mathcal{L}_{\text{LIR}},
\end{equation}
with $\gamma=0.005$ is chosen empirically. AV‐SAL yields robust soft labels; the triplet loss preserves annotated relationships; and LIR corrects false negatives by enforcing learned latent interaction links. We optimize $\mathcal{L}$ via stochastic gradient descent~\cite{ruder2016overview} and backpropagation.

\section{Experiment}
\subsection{Datasets and Evaluations}
We evaluate our model on the audio-visual cross-modal retrieval (AV-CMR) task using VEGAS~\cite{zhou2018visual} and AVE~\cite{tian2018audio} datasets, following standard splits and preprocessing~\cite{zeng2023learning} to ensure proportional category representation. VEGAS contains 28,103 YouTube clips (2–10s) across 10 event categories, with 22,482 for training and 5,621 for testing. AVE has 1,955 clips across 15 categories, with 1,564 for training and 391 for testing. Audio features are extracted as 128-D embeddings per clip via segment-level VGGish~\cite{hershey2017cnn} embeddings averaged over 1-second segments. Visual features are 1024-D descriptors per clip, obtained by average-pooling InceptionV3~\cite{abu2016youtube} frame-level features sampled at 1-second intervals.  

For evaluation, we perform audio-to-visual and visual-to-audio retrieval using cosine similarity between embeddings to rank the retrieved candidates. Mean Average Precision (MAP) is computed for each cross-modal retrieval direction (audio-to-visual and visual-to-audio) and then averaged, following prior works~\cite{zeng2023learning, zeng2022complete}.

\subsection{Implementation Settings}
Each modality branch consists of three 1,024-unit fully connected layers with Tanh activations and 0.15 dropout, followed by a linear projection into an embedding space aligned with the number of event categories (15 for AVE, 10 for VEGAS). We train each dataset separately with a batch size of 400 for 1,000 epochs using triplet loss (margin = 1.2) and the Adam optimizer~\cite{kingma2014adam} (learning rate $1\times10^{-4}$, $\beta_1=0.9$, $\beta_2=0.999$). Training follows a two-stage schedule~\cite{zeng2024anchor}:  
\textbf{1) Pretraining} (Teacher, epochs 1–400): optimize AV-SAL and proxy-based triplet loss on raw embeddings; extract the latent interaction adjacency matrix from soft-label outputs using GRaSP.  
\textbf{2) Latent interaction refinement} (Student, epochs 401–1000) continue triplet and AV-SAL optimization with latent interaction regularization based on the GRaSP graph.  

Experiments are implemented in PyTorch 1.12 on Ubuntu 22.04 with an NVIDIA RTX 3080 GPU (10 GB), using a fixed seed of 42 for reproducibility. Code is available at \footnote{https://github.com/ZenzenDatabase/causalCMR}.

\begin{table}[t]
\vspace{-10pt}
\small
\centering
\caption{\textbf{MAP comparison against state-of-the-art methods.} The best scores are in bold, second-best scores are underlined.}
\begin{NiceTabular}{l|ccc|ccc}
\toprule
\Block{2-1}{\textbf{Models}} &  \Block{1-3}{\textbf{AVE Dataset}} &&  &\Block{1-3}{\textbf{VEGAS Dataset}} \\ \cline{2-7}
                       & A $\rightarrow$ V &  V $\rightarrow$A  & Avg.
                       & A $\rightarrow$ V  &  V $\rightarrow$A  & Avg. \\ \hline
    Random case
    & 0.127 & 0.124 & 0.126
    & 0.110 & 0.109 & 0.109\\
    CCA~\cite{hardoon2004canonical} 
    & 0.190 & 0.189 & 0.190
    & 0.332 & 0.327 & 0.330\\
    KCCA~\cite{akaho2006kernel} 
    & 0.133 & 0.135 & 0.134
    & 0.288 & 0.273 & 0.281\\
    DCCA~\cite{andrew2013deep}  
    & 0.221 & 0.223 & 0.222
    & 0.478 & 0.457 & 0.468\\
    C-CCA~\cite{rasiwasia2014cluster}  
    & 0.228 & 0.226 & 0.227
    & 0.711 & 0.704 & 0.708\\
    C-DCCA~\cite{yu2018category}
    & 0.230 & 0.227 & 0.229
    & 0.722 & 0.716 & 0.719\\
    TNN-C-CCA~\cite{zeng2020deep} 
    & 0.253 & 0.258 & 0.256
    & 0.751 & 0.738 & 0.745\\ 
    VAE-CCA~\cite{zhang2023variational}
    & 0.328 & 0.302 & 0.315
    & 0.821 & 0.824 & 0.822\\ \hline
    CCTL~\cite{zeng2022complete}
    & 0.328 & 0.267 & 0.298
    & 0.766 & 0.765 & 0.766\\
    VideoAdviser~\cite{wang2023videoadviser}
    & - & - & -
    & 0.825 & 0.819 & 0.822\\
    EICS~\cite{zeng2023learning}
    & 0.337 & 0.279 & 0.308
    & 0.797 & 0.779 & 0.788\\
    TLCA~\cite{zeng2023two}
    & 0.410 & 0.451 & 0.431
    & 0.822 & 0.838 & 0.830\\
    MSNSCA~\cite{zhang2023multi}
    & 0.323 & 0.343 & 0.333
    & 0.866 & 0.865 & 0.866\\
    AADML~\cite{zeng2024anchor}
    & \uline{0.890} & \uline{0.883} & \uline{0.887}
    & \uline{0.901} & \uline{0.891} & \uline{0.896}\\
    \hline
    \textbf{Our method}
    & \textbf{0.906} & \textbf{0.899} & \textbf{0.903} (+1.62\%)
    & \textbf{0.913} & \textbf{0.907} & \textbf{0.910} (+1.45\%)\\
\bottomrule
\end{NiceTabular}
\label{table:comparison}
\end{table}
\subsection{Performance Comparison}
We compare our method against representative AV-CMR approaches, including both classical correlation-based techniques and recent deep learning models, grouped into three categories:(1) \textit{CCA-based methods} including CCA~\cite{hardoon2004canonical}, K-CCA~\cite{akaho2006kernel}, Cluster-CCA~\cite{rasiwasia2014cluster}, Deep CCA (DCCA)~\cite{andrew2013deep}, C-DCCA~\cite{yu2018category}, Triplet with Cluster-CCA (TNN-C-CCA)~\cite{zeng2020deep}, and VAE-CCA~\cite{zhang2023variational}. These methods learn projections (linear or non-linear) to maximize inter-modal correlation. (2) Representation learning with transformer or distillation like MSNSCA~\cite{zhang2023multi} employs a multi-scale network with shared cross-attention for fine-grained audio–visual alignment. VideoAdviser~\cite{wang2023videoadviser} method utilizes CLIP-based teacher–student distillation for robust audio–visual embeddings. (3) \textit{Metric learning methods} include CCTL~\cite{zeng2022complete}, EICS~\cite{zeng2023learning}, and AADML~\cite{zeng2024anchor}. These approaches learn discriminative embeddings via contrastive or triplet objectives.

As shown in Table~\ref{table:comparison}, our model achieves the highest MAP on both datasets, outperforming the strongest baseline by 1.62\% on AVE and 1.45\% on VEGAS. In cross-modal retrieval, such gains are notable given the strength and near-saturation of existing baselines, underscoring the effectiveness of our approach.

\subsection{Ablation Study}
\label{sec:ablation}
\subsubsection{Latent interaction graph analysis}
We track how cross-modal latent interactions evolve by applying GRaSP to soft-label outputs at seven checkpoints (epochs=300, 400, …, 900). We summarize temporal consistency with a frequency matrix \(\mathbf{F}\in[0,1]^{2C\times 2C}\):

\begin{equation} 
\small
F_{ij} \;=\; \frac{1}{T}\sum_{t=1}^T \mathbf{1}\{\widehat{A}^{(t)}_{ij} > \varepsilon\}, \qquad T=7,\; \varepsilon\ge0,
\end{equation}
where \(\mathbf{F}_{ij}\) is the fraction of checkpoints where edge \((i\!\to\!j)\) appears.

\begin{figure}[t]
  \centering
  \includegraphics[width=0.9\linewidth]{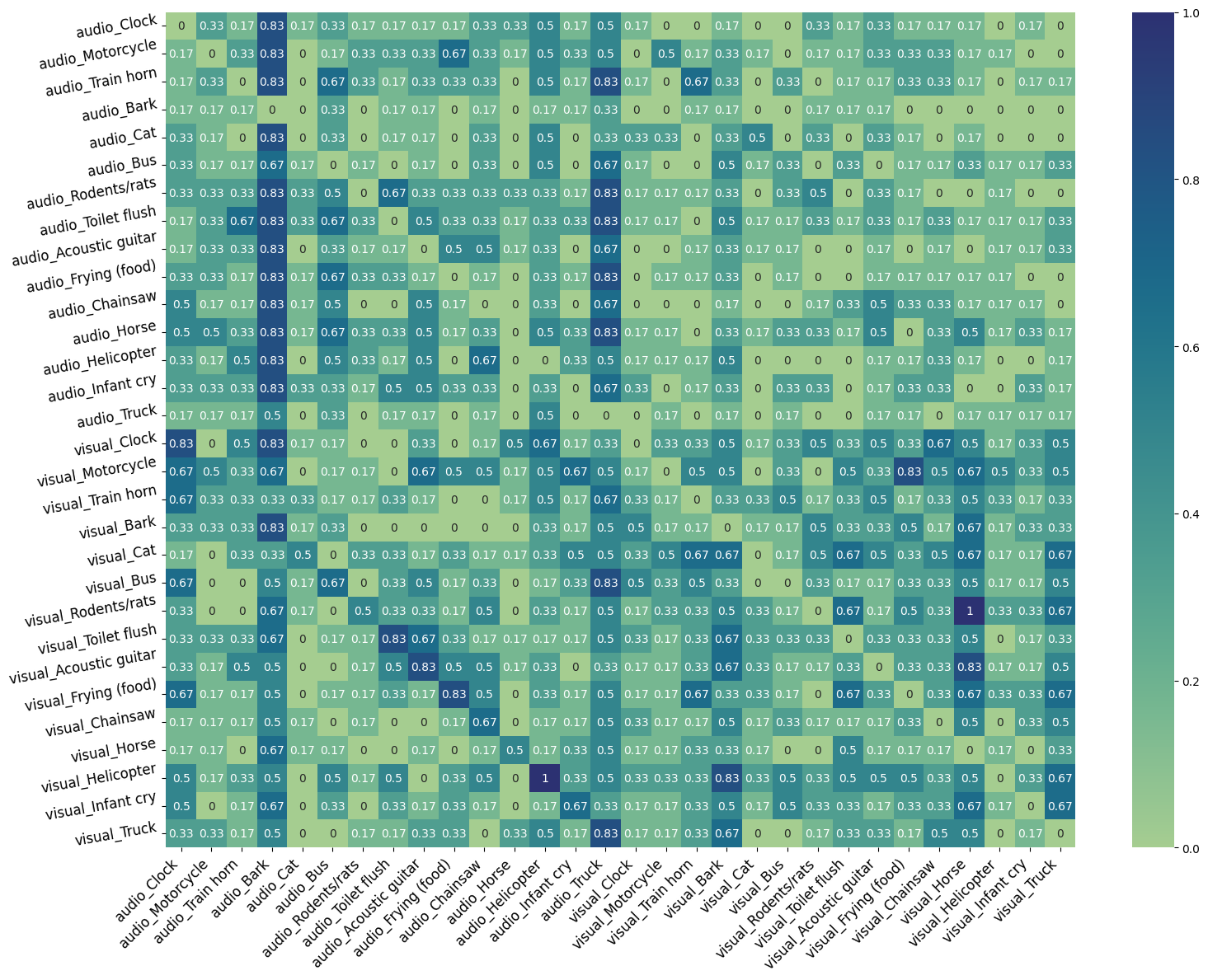}
  \caption{\textbf{Latent interaction edge frequency heatmap} (on AVE). Each cell shows how often class \(i\) was inferred to influence class \(j\) across seven checkpoints.}
  \label{fig:causal_heatmap}
  \vspace{-10pt}
\end{figure}

Figure~\ref{fig:causal_heatmap} visualizes $\mathbf{F}$. High-frequency cross-modal links (e.g., ``Helicopter (visual) $\rightarrow$ Helicopter (audio)'' 0.99; ``Truck (visual) $\rightarrow$ Truck (audio)'' 0.83; ``Frying (visual) $\rightarrow$ Frying (audio)'' 0.83) indicate persistent, robust associations learned across training. In total, 81 class-pairs exceed a frequency of 0.714 (i.e., appear in at least 5 of 7 checkpoints), showing that certain latent interaction links are consistently recovered—likely reflecting strong temporal/contextual coupling rather than spurious co-occurrence. These stable patterns underscore the utility of latent interaction inferred for revealing high-confidence cross-modal relationships.

\begin{figure}[t]
  \centering
  \includegraphics[width=1.05\linewidth]{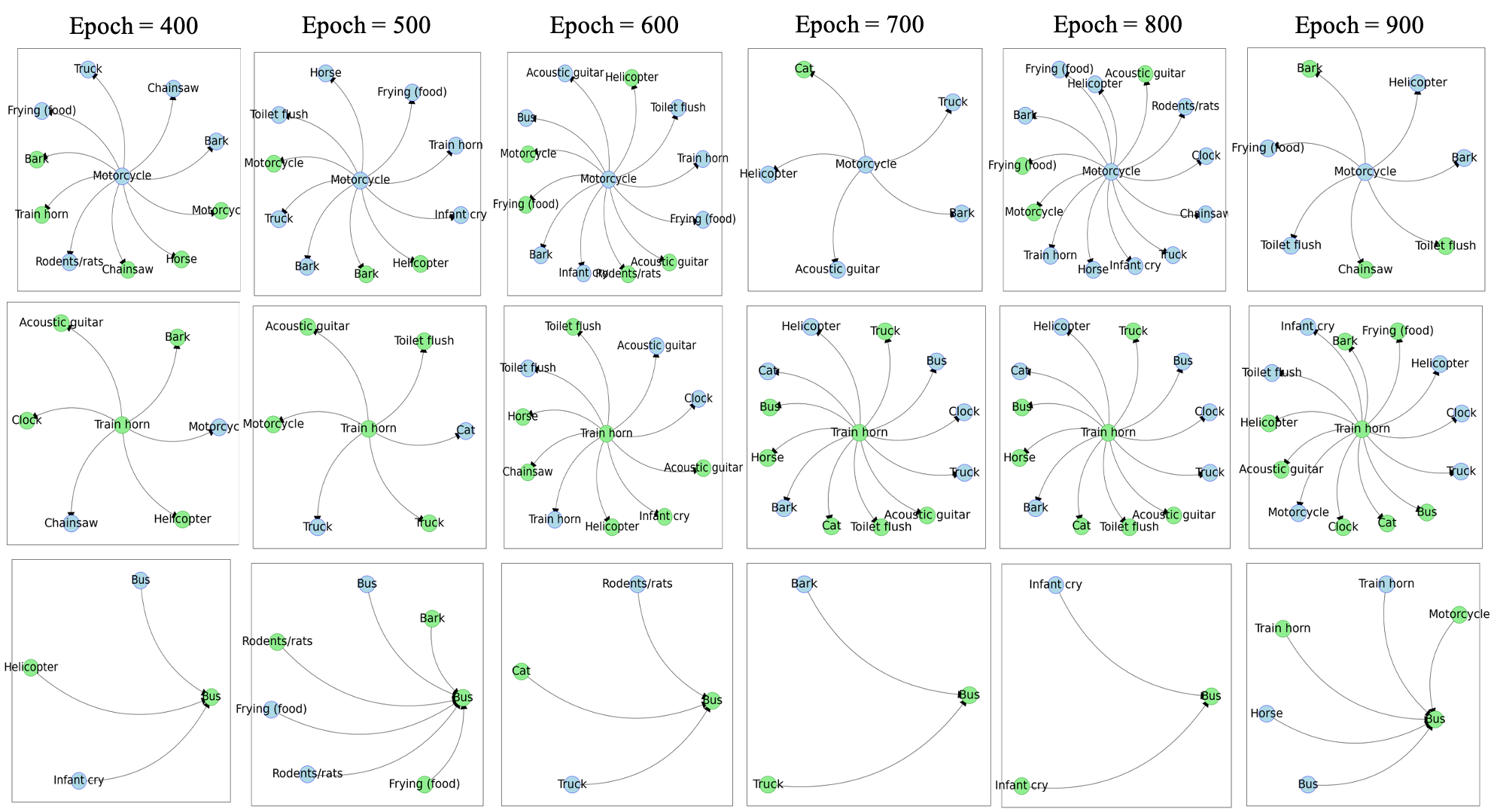}
   \caption{Latent interaction subgraphs from the AVE dataset across training epochs (400–900). The top row shows subgraphs centered on “Motorcycle (audio)” as the cause, the middle row on “Train horn (visual)” as the cause, and the bottom row on “Bus (visual)” as the effect. Blue nodes denote audio classes, and green nodes denote visual classes.}
  \label{fig:example_causal}
  \vspace{-10pt}
\end{figure}

\textbf{Case studies of evolving latent interaction subgraphs.}  
Figure~\ref{fig:example_causal} shows three representative subgraphs across training epochs (top: ``Motorcycle (audio)'' as source; middle: ``Train horn (visual)'' as source; bottom: ``Bus (visual)'' as target).  
\textbf{(i) ``Motorcycle (audio)'' (source).} High activation of this node is consistently associated with increased model activations for audio classes (e.g., ``Bark'', ``Chainsaw'') and visual classes (e.g., ``Motorcycle'', ``Train horn'', ``Helicopter''). Stable links to mechanically related events suggest engine-like sounds reliably indicate outdoor/mechanical scenes, while some connections (e.g., to ``Helicopter'' or ``Toilet flush'') fluctuate across epochs.  
\textbf{(ii) ``Train horn (visual)'' (source).} Visual train-horn cues repeatedly predict audio events (e.g., ``Clock'', ``Motorcycle'', ``Helicopter'') and nearby visual classes (e.g., ``Truck'', ``Toilet flush''), reflecting acoustically rich transportation contexts.  
\textbf{(iii) ``Bus (visual)'' (target).} As a target node, ``Bus (visual)'' receives directed inputs from diverse sources (``Infant cry'', ``Frying (food)'', ``Rodents/rats'', ``Train horn''), indicating bus scenes occur in varied urban, domestic, and transit contexts.  

Together, these examples show the inferred directed dependencies are interpretable yet dynamic: some associations are consistently recovered (high-confidence signals), while others evolve during training—supporting the use of model-inferred dependency information to improve cross-modal embedding learning.
\begin{figure}[b]
\vspace{-20pt}
  \centering
  \includegraphics[width=0.9\linewidth]{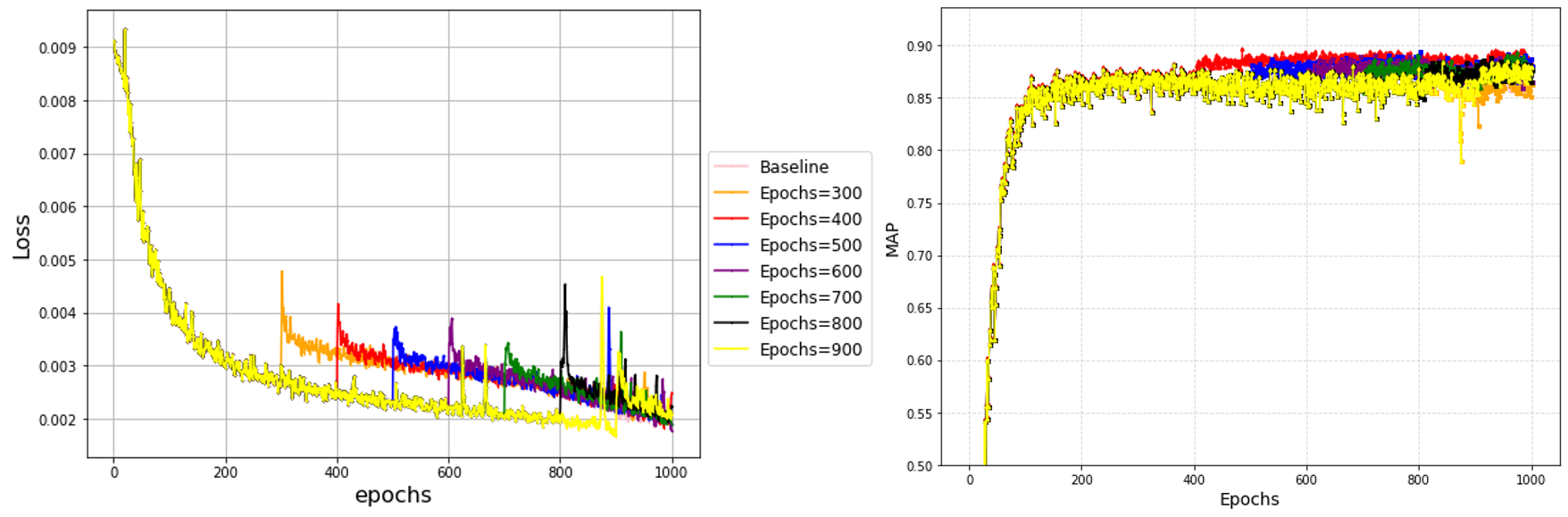}
    \caption{Training loss (left) and test MAP (right) on AVE for different latent-interaction insertion epochs \(M\in\{300,400,\dots,900\}\) (baseline = no insertion).}
  \label{fig:loss_map}
\end{figure}
\subsubsection{Impact of the inferred latent interaction}
We study sensitivity to when the inferred latent interaction graph is injected by varying the insertion epoch \(M\in\{300,400,500,600,700,800,900\}\) (baseline: no insertion). Figure~\ref{fig:loss_map} plots training loss and test MAP on AVE for these settings. Introducing the inferred interaction at \(M\) produces an immediate perturbation in the loss followed by rapid stabilization, showing the optimizer adapts to the additional regularization. Compared to the baseline (no interaction), all interaction-aware settings improve MAP, indicating the inferred dependencies provide useful supervision. The best trade-off is obtained at \(M=400\): loss converges smoothly and test MAP is highest. Introducing interaction too early can disturb initial representation learning, while introducing it too late reduces its ability to shape embeddings; \(M=400\) balances label reliability for graph inference with sufficient remaining training time for the graph to guide refinement.  

\begin{table}[t!]
\small
    \centering
    \caption{Component-wise impact on the final objective loss for different metric learning methods on the AVE and VEGAS datasets.}
    \begin{NiceTabular}{l|ccc|ccc}
        \toprule
        \multicolumn{1}{c}{\textbf{Models}} & \multicolumn{3}{c|}{\textbf{AVE dataset}} & \multicolumn{3}{c}{\textbf{VEGAS dataset}} \\
        \cmidrule(lr){2-4} \cmidrule(lr){5-7}
        \textbf{ } & A$\rightarrow$V & V$\rightarrow$A & Avg. & A$\rightarrow$V & V$\rightarrow$A & Avg. \\
        \midrule
        \textbf{AV-SAL+Proxy (Triplet)+LIR} 
        & \textbf{0.906} & \textbf{0.899} & \textbf{0.903}
        & \textbf{0.913} & \textbf{0.907} & \textbf{0.910}  \\
        AV-SAL+Proxy (Hard Triplet)+LIR & 0.893 & 0.889 & 0.891 & 0.873 & 0.867 & 0.870 \\
        AV-SAL+Proxy (Contrastive)+LIR & 0.822 & 0.858 & 0.840 & 0.795 & 0.814 & 0.805 \\
        AV-SAL+Proxy (N-pair)+LIR & 0.835 & 0.846 & 0.841 & 0.658 & 0.611 & 0.634 \\
        \midrule
        AV-SAL+Proxy (Triplet) & \underline{0.890} & \underline{0.883} & \underline{0.887} & \underline{0.901} & \underline{0.890} & \underline{0.896} \\
        AV-SAL+Proxy (Hard Triplet) & 0.883 & 0.885 & 0.884 & 0.865 & 0.851 & 0.858 \\
        AV-SAL+Proxy (Contrastive) & 0.798 & 0.854 & 0.826 & 0.783 & 0.807 & 0.795 \\
        AV-SAL+Proxy (N-pair) & 0.828 & 0.833 & 0.830 & 0.647 & 0.612 & 0.629 \\
        \midrule
        AV-SAL+Triplet+LIR & 0.408 & 0.451 & 0.430 & 0.763 & 0.727 & 0.745 \\
        AV-SAL+Hard Triplet+LIR & 0.407 & 0.439 & 0.423 & 0.774 & 0.759 & 0.767 \\
        AV-SAL+Contrastive+LIR & 0.376 & 0.381 & 0.379 & 0.712 & 0.714 & 0.713 \\
        AV-SAL+N-pair+LIR & 0.326 & 0.340 & 0.333 & 0.568 & 0.550 & 0.559 \\
        \midrule
        Proxy (Triplet\textsuperscript{\cite{schroff2015facenet}})+LIR &0.886 &0.880 &0.883 &0.890 &0.883 &0.887 \\
        Proxy (Hard Triplet\textsuperscript{\cite{schroff2015facenet}})+LIR & 0.877 & 0.880 & 0.879 & 0.857 &0.846 &0.852 \\
        Proxy (Contrastive\textsuperscript{\cite{hadsell2006dimensionality}})+LIR & 0.790 & 0.844 & 0.817 & 0.776 &0.794 &0.785 \\
        Proxy (N-pair\textsuperscript{\cite{sohn2016improved}})+LIR & 0.822 & 0.828 & 0.825 & 0.642 &0.603 &0.623 \\
        \bottomrule
    \end{NiceTabular}
    \vspace{-10pt}
    \label{table:loss_map_ave}
\end{table}

\subsubsection{Impact of loss components} 
Table~\ref{table:loss_map_ave} reports MAP scores on AVE and VEGAS for different combinations of AV-SAL, proxy-based metric learning (Triplet, Hard Triplet, Contrastive, N-pair), and LIR. The table blocks isolate the contribution of each component: (1) AV-SAL + Proxy (+Loss) + LIR (full model), (2) AV-SAL + Proxy (+Loss) without LIR, (3) AV-SAL + Loss + LIR (no Proxy), and (4) Proxy (+Loss) + LIR (no AV-SAL). The full model performs best: AV-SAL + Proxy (Triplet) + LIR achieves the highest MAP (\textbf{0.903} on AVE, \textbf{0.910} on VEGAS), outperforming the closest AV-SAL+Proxy baseline by +0.016 and +0.014, respectively. 

This demonstrates that LIR sharpens cross-modal alignment, especially when paired with AV-SAL and Proxy. Among proxy-based losses, triplet supervision is the strongest, yielding the most robust gains, producing the most discriminative embeddings.  Overall, the ablation confirms that \emph{AV-SAL + Proxy (Triplet) + LIR} is the optimal configuration, with AV-SAL and LIR providing complementary benefits.

\begin{figure}[t]
\centering
\includegraphics[width=0.9\textwidth]{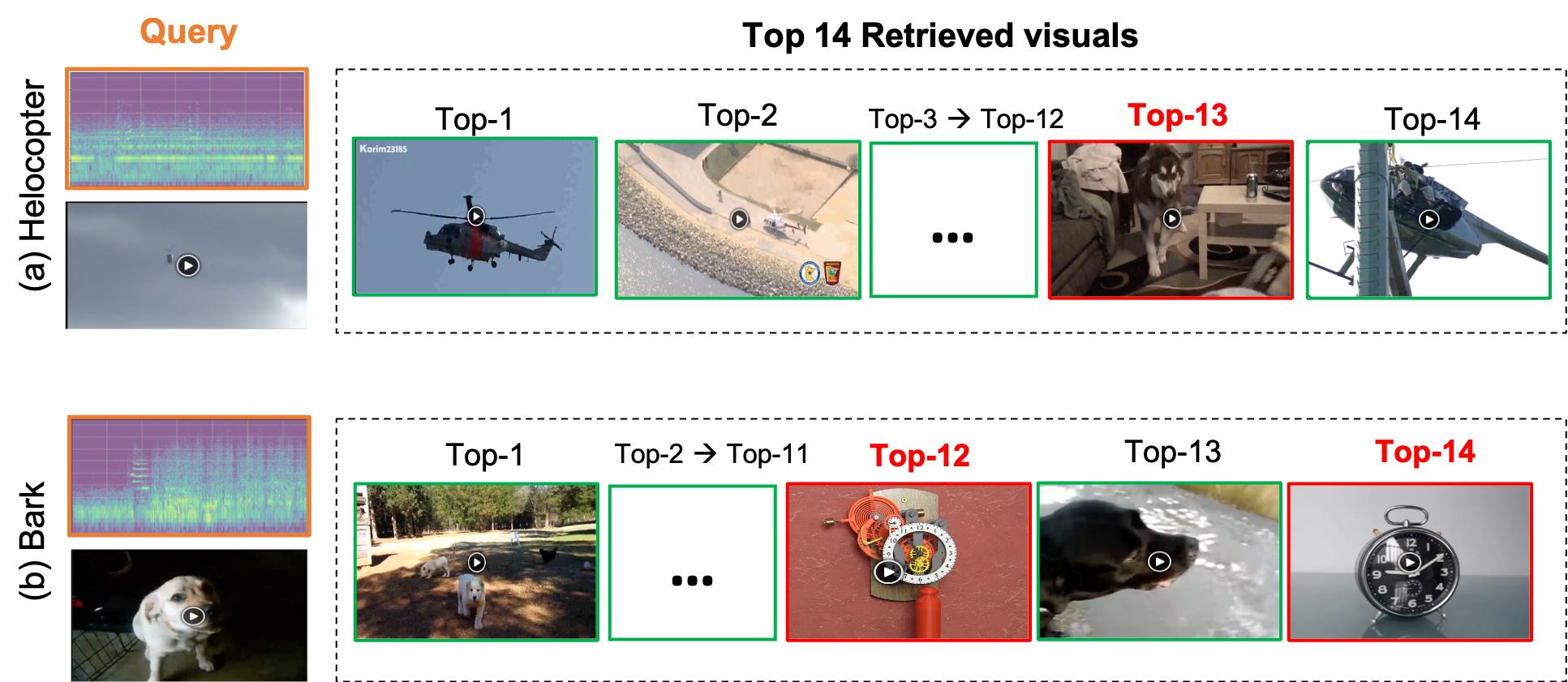}
\caption{Qualitative audio-to-visual retrieval on AVE. For two audio queries with different labels, the top-14 retrieved visuals are shown. Correct matches are highlighted in green, incorrect ones in red.}
\label{fig:retrieval_case}
\vspace{-10pt}
\end{figure}
\subsection{Qualitative Results}
We present qualitative audio-to-visual retrieval examples on the AVE dataset to assess the semantic alignment of the learned cross-modal embeddings (Figure~\ref{fig:retrieval_case}). 
The few retrieval errors can be explained by the learned interaction strengths in the inferred dependency graph (Fig. \ref{fig:causal_heatmap}). For the Helicopter audio query, the model strongly associates audio–helicopter with visual–helicopter (edge frequency = 1.0), leading to predominantly correct retrievals. The single incorrect Dog bark visual at rank 13 corresponds to a weaker but non-zero dependency (audio–helicopter → visual–dog bark, frequency = 0.5), indicating partial confusion likely due to shared acoustic characteristics (e.g., high-energy, outdoor sounds). In contrast, for the Dog bark audio query, there is no inferred dependency between audio–dog bark and visual–clock (frequency = 0), nor a strong directional dependency to visual–dog bark. As a result, the few incorrect Clock samples at lower ranks reflect residual ambiguity not reinforced by the interaction graph. Overall, correct retrievals align with stronger self-dependencies, while errors occur at low ranks where dependency signals are weak or absent, indicating that the graph regularization influences—but does not rigidly dictate—the embedding structure.

\section{Conclusion}
We present a framework for audio–visual embedding learning that addresses the challenges of sparse annotations and incidental co-occurrences. Our approach first aligns teacher-generated soft labels across modalities (AV-SAL), then applies the GRaSP algorithm to infer a directed latent interaction graph, and finally enforces these links through Latent Interaction Regularization (LIR), effectively disentangling meaningful semantic dependencies from superficial correlations. Experiments on AVE and VEGAS show consistent MAP improvements of approximately 1.5\% over state-of-the-art baselines, demonstrating more robust and semantically coherent embeddings. In future work, we plan to explore richer latent interaction modeling—potentially capturing principles akin to causality, and reduce reliance on manual labels by incorporating self-supervised strategies.

\section{Discussion and Clarification of Review Concerns}
\textbf{ILI graph - not a causal claim}. We do not claim that the inferred ILI graph is definitive causal evidence; rather, GRaSP is applied to the teacher logits to recover an estimated, sparse, directed dependency matrix from observational model outputs. The directed edges, therefore, represent asymmetric predictive dependencies in the teacher’s logit/representation space (how activation of class $i$ helps predict class $j$), not proven interventions or causal mechanisms. To reduce spurious links, we use diagonal-zeroing and $L1$ normalization, thresholding and edge-stability aggregation across checkpoints, and we restrict LIR to high-weight, stable edges, all of which bias the graph toward compact, reproducible dependencies. Importantly, the graph is used only as a soft regularizer (LIR) alongside AV-SAL and triplet supervision, so it nudges embeddings toward consistent co-activations without enforcing hard, possibly incorrect relations. Finally, we acknowledge that some edges will reflect acoustic similarity or frequent co-occurrence rather than semantic causation; establishing true causality would require interventions or experimental validation and is beyond the scope of this observational study.

\textbf{Scalability and graph sparsity}. GRaSP’s search does not scale trivially to very large class vocabularies, which motivates our experiments on small label sets (10, 15 classes). In practice, GRaSP is an offline, infrequent training-time step, and GRaSP’s sparsity-inducing objective produces compact graphs with a modest, fixed $lamba$ (we do not heavily tune $lamba$). For larger label sets, practical adaptations can be used, e.g., infer dependencies on clustered/meta-classes, restrict inference to top-K/high-confidence labels per sample, or adopt approximate/greedy/sampling-based search or parametric dependency estimators—to trade a small approximation error for tractability. Importantly, these adaptations preserve the deployment simplicity: only the student model is used at inference.

\textbf{Dependency on teacher quality}: We acknowledge the risk of propagating teacher errors. To mitigate this, the teacher is trained with hard labels plus progressive self-distillation so its outputs are grounded and better calibrated; the ILI graph is inferred from aggregated soft-label distributions and thresholded to produce a sparse graph that discards weak links; and LIR is applied as a soft regularizer alongside the supervised loss rather than as hard supervision. These design choices dampen spurious teacher signals rather than enforce them. Empirical stability of inferred edges across checkpoints further reduces sensitivity to occasional teacher noise.

\textbf{Pipeline complexity}: Our method uses a coordinated two-stage schedule, not two different model architectures. Both teacher and student share the same modality encoders and embedding architecture. The teacher stage runs from epoch 1 to the transition epoch $M$ (e.g., $M \in$ \{300,400,…,900\}) to learn calibrated soft labels; at $M$, we infer the sparse ILI graph once; the student stage then continues from 
$M$ to 1000 with the Latent Interaction Regularizer (LIR) added to the loss. Thus the extra complexity is a training-time schedule choice rather than a multi-model deployment cost: at inference only the compact student is used (no teacher or graph), so runtime latency, memory, and engineering integration remain equivalent to standard single-model baselines. Our ablations (Sec. \ref{sec:ablation}, Fig. \ref{fig:loss_map}) show that a single, well-timed graph insertion yields consistent gains while keeping added training work bounded.
\bibliographystyle{plain}
\bibliography{mybib}
\end{document}